\documentclass[12pt]{article}
\usepackage{amsmath,amssymb,amsfonts}

\begin{document}

\title{Relativistic accelerating electromagnetic waves}

\author{Shahen Hacyan
}

\renewcommand{\theequation}{\arabic{section}.\arabic{equation}}

\maketitle
\begin{center}

{\it  Instituto de F\'{\i}sica,} {\it Universidad Nacional Aut\'onoma de M\'exico,}

{\it Apdo. Postal 20-364, M\'exico D. F. 01000, Mexico.}

e-mail: hacyan@fisica.unam.mx

\end{center}
\vskip0.5cm

\begin{abstract}

An exact solution of the Maxwell equations in Rindler coordinates is obtained. The electromagnetic field represents
a wave preserving its shape in a relativistic uniformly accelerated frame. The relation with Airy beams is
shown explicitly in the non-relativistic limit.

\end{abstract}

PACS: 41.20.Jb;42.25.Bs


 \maketitle

\newpage
\section{Introduction}

In 1979, Berry and Balazs \cite{BB} obtained a solution of the Schr\"{o}dinger equation, in terms of an Airy
function, describing a non-spreading accelerating wave packet. More recently, the Airy beam, an equivalent
solution of the paraxial equation of optics, has received a considerable amount of attention both from a theoretical
as well as from an experimental point of view (see, e.g., Ref. \cite{prl, 2010a, 2010b}).

These previous works on accelerating waves have been restricted to the non-relativistic case. The aim of the
present work is to obtain an exact solution of the Maxwell equations in Rindler coordinates, which are the natural
coordinates associated with uniform acceleration in special relativity. The ensuing exact solution represents an
electromagnetic wave that is shape invariant in a uniformly accelerated frame and appears therefore
 as an accelerating wave in an inertial frame. The relation with the Airy beam is shown explicitly in the
non-relativistic limit.

In Section 2, the Maxwell equations are solved in Rindler coordinates for the two fundamental modes of the field.
A particular solution, invariant under space-time hyperbolic rotations, is obtained and the explicit forms of the
energy density and the Poynting vector are worked out. The relation with the Airy beam in the non-relativistic
limit is shown in Section 3 and a brief discussion of the results is given in Section 4.

\section{Maxwell equations}
\setcounter{equation}{0}

The Rinder metric is obtained with a change from Cartesian $(ct,x,y,z)$ to Rindler coordinates $(cT, x, y, Z)$ defined as
$$
ct= Z \sinh (a T),
$$
\begin{equation}
z= Z \cosh (a T),\label{zt}
\end{equation}
where $a$ is a parameter with dimensions of frequency (or acceleration divided by $c$). The metric thus takes
the form
\begin{equation}
ds^2 = -a^2 Z^2 dT^2 + dx^2 + dy^2 + dZ^2.
\end{equation}

The electromagnetic field tensor $F_{\alpha \beta}$ follows from a vector potential $A_{\alpha}$ such that
\begin{equation}
F_{\alpha \beta} = \partial_{\alpha} A_{\beta} - \partial_{\beta} A_{\alpha} \label{3}
\end{equation}
and is assumed to satisfy the source-free Maxwell equations
\begin{equation}
\frac{\partial}{\partial x^{\mu}} \Big( Z F^{\alpha \mu} \Big)= 0 .\label{2}
\end{equation}
The two modes of the electromagnetic field in Rindler coordinates can be represented by the
transverse electric (E) and transverse magnetic (M) vector potentials,
\begin{equation}
A_{\mu}^E = (0;~\partial_2 U^E ,~ -\partial_1 U^E ,~ 0)
\end{equation}
and
\begin{equation}
A_{\mu}^M = (aZ \partial_3 U^M;~ 0,~0 ,~ \frac{1}{aZ} \partial_0 U^M)~,
\end{equation}
given in terms of two potentials $U^E$ and $U^M$.
The total electromagnetic field turns out to be
\begin{eqnarray}
  &F_{12}& = -\partial^2_{11} U^E- \partial^2_{22} U^E  ~,\\ \nonumber
  &F_{23}& = \frac{1}{aZ} \partial^2_{20} U^M + \partial^2_{13} U^E ~,\\ \nonumber
  &F_{31}&= ~-~\frac{1}{aZ} \partial^2_{10} U^M +\partial^2_{23} U^E ~,\\ \nonumber
& F_{01}& =- aZ \partial^2_{13} U^M +\partial^2_{20} U^E ~,\\ \nonumber
& F_{02}&=- aZ \partial^2_{23} U^M-\partial^2_{10} U^E  ~,\\ \nonumber
& F_{03}&=- \partial_3 \Big(aZ \partial_3 U^M\Big) + \frac{1}{aZ} \partial^2_{00} U^M .
\end{eqnarray}

As can be checked by direct substitution, the complete Maxwell equations are satisfied if
\begin{equation}
\partial^2_{00} U - a^2 Z^2 ~\nabla^2_{\bot} U -
a^2 Z~\partial_3 \Big(Z~ \partial_3 U \Big) =0~,\label{pot}
\end{equation}
for both potential $U^E$ and $U^M$. Here $\nabla^2_{\bot} = \partial^2_{11} + \partial^2_{22}$.

This last equation admits a solution by separation of variables:
\begin{equation}
U(x^{\mu}) = e^{-i \Omega T + i k_x x + i k_y y} U(Z),
\end{equation}
where $\Omega$ is the frequency as measured in the accelerated frame and $U(Z)$ satisfies the equation
\begin{equation}
Z^2 \frac{d^2}{dZ^2} U + Z \frac{d}{dZ} U + \Big[ \Big( \frac{\Omega }{a} \Big)^2- k_{\bot}^2 Z^2 \Big] U
=0,\label{EQM}
\end{equation}
for both modes, with $k^2_{\bot} \equiv k_x^2 + k_y^2$. The solution is
\begin{equation}
U (Z)= C_{(E,M)} K_{i \Omega /a} (k_{\bot}Z),
\end{equation}
where $K_{i \Omega /a} (k_{\bot}Z)$ is the modified Bessel function of the third kind with imaginary order and
$C_{(E,M)}$ are constants. The other linearly independent solution is $L_{i \Omega /a} (k_{\bot}Z)$, but it
diverges at $Z \rightarrow \infty$ \cite{dun}.

The electromagnetic tensor has the following forms in Rindler coordinates. For the transverse electric mode:
$$
F_E^{\alpha \beta} = C_E e^{-i \Omega T + i k_x x + i k_y y}
$$
\begin{equation}
\times \frac{c^2}{(aZ)^2} \begin{pmatrix}
  0 &  -c^{-1} \Omega k_y K & c^{-1}\Omega k_x K &  0 \\
   c^{-1}\Omega k_y K  & 0 & k^2_{\bot}(aZ)^2 K  & -ik_y k_{\bot}(aZ)^2 K' \\
   -c^{-1}\Omega k_x K  & -k^2_{\bot}(aZ)^2 K & 0 & ik_x k_{\bot}(aZ)^2 K' \\
   0  & ik_y k_{\bot}(aZ)^2 K' & -ik_x k_{\bot}(aZ)^2 K' & 0
\end{pmatrix},
\end{equation}
and for the transverse magnetic mode:
$$
F_M^{\alpha \beta} = C_M e^{-i \Omega T + i k_x x + i k_y y}
$$
\begin{equation}
\times \frac{c}{aZ} \begin{pmatrix}
  0 &  -ik_x k_{\bot} K' & -i k_y k_{\bot} K' &  -k^2_{\bot}K \\
   ik_x k_{\bot} K' & 0 & 0 & -c^{-1} \Omega k_x  K \\
   ik_y k_{\bot} K'  & 0 & 0 & -c^{-1}\Omega k_y K \\
   k^2_{\bot} K  & c^{-1}\Omega k_xK & c^{-1}\Omega k_yK & 0
\end{pmatrix},
\end{equation}
where $K \equiv K_{i \Omega /a} (k_{\bot}Z)$ and $K'$ is the derivative of $K$ with respect to its argument.

\subsection{Minkowski coordinates}

Transforming to Minkowski coordinates (the electric and magnetic fields are $E_a = F_{a0}$ and $B_a =
\frac{1}{2}~ \epsilon_{abc} F^{bc}$), it follows that
\begin{equation}
{\bf E}=  e^{-i \Omega T + i k_x x + i k_y y} \Big[-C_E G ~{\bf k}_{\bot} \times {\bf e_z}+ C_M \Big(F ~{\bf
k}_{\bot} - k_{\bot}^2 K ~{\bf e_z}\Big) \Big]
\end{equation}
\begin{equation}
{\bf B} =  e^{-i \Omega T + i k_x x + i k_y y}  \Big[-C_E \Big( F ~{\bf k}_{\bot} - k_{\bot}^2~ {\bf e_z} \Big)
-C_M G ~{\bf k}_{\bot} \times {\bf e_z}  \Big],
\end{equation}
where
\begin{equation}
F \equiv  \frac{\Omega}{aZ}   \sinh (aT) K - i k_{\bot}\cosh (aT)  K'
\end{equation}
\begin{equation}
G \equiv \frac{\Omega}{aZ} \cosh (aT)  K - i  k_{\bot}\sinh (aT) K'.
\end{equation}
In these formulas, $T$ and $Z$ must be interpreted as functions of $t$ and $z$, as given by Eq. (\ref{zt}).

It follows from these equations that the invariants of the electromagnetic field are
\begin{equation}
{\bf E} \cdot {\bf B}^* + {\rm c.c.}  = -(C_E C_M^* + C_M C_E^*) k_{\bot}^4 {\cal F}(k_{\bot} Z)
\end{equation}
and
\begin{equation}
{\bf E} \cdot {\bf E}^* - {\bf B} \cdot {\bf B}^*  = (|C_M|^2 - |C_E|^2)  k_{\bot}^4 {\cal F}(k_{\bot} Z),
\end{equation}
where
\begin{equation}
{\cal F}(k_{\bot} Z) \equiv   (K')^2+ \Big[1 - \Big( \frac{\Omega }{ak_{\bot} Z} \Big)^2 \Big] K^2 ~.
\end{equation}

As for the energy density $W$ and Poynting vector ${\bf S}$, it follows that
\begin{equation}
W={\bf E} \cdot {\bf E}^* + {\bf B} \cdot {\bf B}^*  = (|C_E|^2 +|C_M|^2) k_{\bot}^4 \Big[ \cosh (2 a T )~ {\cal
G}(k_{\bot} Z) + K^2\Big]
\end{equation}
and
$$
{\bf S} = {\bf E} \times {\bf B}^* + {\rm c.c.} = k_{\bot}^4 \cosh (aT) \times
$$
$$
\Big\{ (|C_E|^2 +|C_M|)^2  \Big[ \sinh (aT)~{\cal G}(k_{\bot} Z) ~{\bf e}_z +2 \Big( \frac{\Omega }{ak_{\bot} Z}
\Big) K^2 {\bf e}_{\bot} \Big] +
$$
\begin{equation}
2i (C_M C_E^* -C_E C_M^*) K K' ~{\bf e}_{\bot} \times {\bf e}_z \Big\} ,\label{poyM}
\end{equation}
where
\begin{equation}
{\cal G}(k_{\bot} Z) \equiv  (K')^2   + \Big( \frac{\Omega }{ak_{\bot} Z} \Big)^2 \label{G}
\end{equation}
and ${\bf e}_{\bot} = {\bf k}_{\bot} /k_{\bot}$ is the unit vector in the direction of ${\bf k}_{\bot}$.

The total energy integrated over the whole Rindler wedge, $z > 0$, is unbounded due to the divergence of the
modified Bessel function at the horizon $z=\pm ct$. This is related to the nonphysical behavior
of the field in that limit, since it would correspond to a section of the electromagnetic field with infinite
acceleration. Nevertheless, it is possible to integrate from a certain value of $Z$, say $\epsilon$, to $Z
\rightarrow \infty$ and obtain a finite value of the energy. According to Eq (\ref{I}) in the Appendix,
\begin{equation}
\int_{\epsilon}^{\infty} W ~dz =  (|C_E|^2 +|C_M|^2)  \frac{\pi k_{\bot}^4 \Omega }{ \epsilon a  \sinh (\pi \Omega /a)} ~ ,
\end{equation}
the integration being taken over the three-dimensional space $t=0=T$ and $z > 0$.

\section{Non-relativistic limit}
\setcounter{equation}{0}

The above solutions are exact. The non-relativistic limit can be obtained defining first an acceleration parameter
$g=a c$ and then performing a coordinate shift
$$
z \rightarrow z + \frac{c^2}{g},
$$
from where it follows that
\begin{equation}
Z  \approx \frac{c^2}{g} + z - \frac{1}{2} g t^2 \label{nr}
\end{equation}
and
\begin{equation}
\sinh  (aT ) \approx \frac{g}{c}t \quad , \quad \cosh ( aT ) = 1 + O[(gt/c)^2],
\end{equation}
and of course $T \approx t$, thus identifying $\Omega$ as the frequency in an inertial frame in which the velocity
of the wave is non relativistic. As a next step, one can use some approximation for the Bessel function of
imaginary order (see, e.g, \cite{dun,abr,bal}). However, as shown by Balogh \cite{bal}, such approximations break down for values of the argument close to the order (the so-called turning point) and another kind of approximation must be used involving
Airy functions. For our present purpose it is enough to notice that, in the
non-relativistic approximation of Eq. (\ref{nr}), Eq (\ref{EQM}) reduces to
\begin{equation}
\frac{d^2 U}{ d\zeta^2} + \Big[ \Big( \frac{\Omega}{c } \Big)^2 -k_{\bot}^2 \Big(1 + 2 \frac{g}{c^2}\zeta \Big)
\Big] U=0,
\end{equation}
where $\zeta \equiv z - \frac{1}{2} a t^2$. The solution in terms of the Airy function $Ai$ is
\begin{equation}
U= C_{E, M} ~Ai \Big[ k_{\bot}^{1/3} \Big( \frac{2g}{c^2} \Big)^{2/3}\zeta + \frac{1}{2g k_{\bot}} (k_{\bot}^2 c^2
- \Omega^2) \Big].
\end{equation}
The particular case $k_{\bot} c = \Omega$ corresponds to the turning point \cite{bal} where the solution in terms
of Bessel functions reduces to the Airy beam treated in the literature. To obtain the non-relativistic limit, it
is only necessary to change the function $K$ of the previous section to the Airy function given above, but this
does not seem to offer a significant advantage for practical calculations.

\section{Discussion}

An exact solution of Maxwell's equations in Rindler coordinates was obtained. From the expression of the Poynting
vector in the inertial frame, given by Eq. (\ref{poyM}), it is seen that it describes an accelerating light beam
along the $z$ axis with a turning point at time $t=0$. It also has a component perpendicular to the averaged
direction of propagation, which is typical of structured light beams (for instance, Bessel beams \cite{hj}).

The above solution is given in terms of a modified Bessel function of imaginary order and is a linear
superposition of the two modes of the electromagnetic field. In the non-relativistic limit, it reduces to an Airy
wave for a particular value of the frequency. It should be noticed, however, that the total energy of the
relativistic accelerating field is not bounded due to its divergence at the Rindler horizon $z=\pm ct$; this
suggests that the Airy wave, if properly extended to the exact relativistic limit, will exhibit a similar
divergence. This fact is not unexpected since the energy of an electromagnetic field integrated over the whole
space is usually divergent in many idealized situations, including plane waves, and it is necessary to constrain
the field to a finite region of space. Moreover, as shown by Boulware \cite{bou}, singularities at the horizon
also occur in the related problem of the electromagnetic field produced by a uniformly accelerated charge.

\section*{Appendix}

\setcounter{section}{0}

\setcounter{equation}{0}

The modified Bessel function of the third kind is defined in general as
\begin{equation}
K_{\mu} (x) = \int_0^{\infty} e^{-x \cosh u} \cosh (\mu u) ~du.
\end{equation}
For purely imaginary order, $K_{i\alpha}(x)$ has the following asymptotic behaviors (see, e.g. Dunster \cite{dun}): near $x=0$
\begin{equation}
K_{i\alpha}(x) = - \Big( \frac{\pi}{\alpha \sinh (\alpha \pi)} \Big)^{1/2} \{ \sin [\alpha \ln (x/2) - {\rm Arg}
(\Gamma (1+i\alpha))] + O(x^2) \},
\end{equation}
and for $x \rightarrow \infty$
\begin{equation}
K_{i\alpha}(x) =  \Big( \frac{\pi}{2x}\Big)^{1/2} e^{-x} \{1+ O(1/x) \}.
\end{equation}

The energy-density is proportional to the function ${\cal E}(x) \equiv {\cal G}(x)+ K_{i\alpha}^2(x)$, where ${\cal G}$ is defined by (\ref{G}).
The recurrence relation
$$
K'_{i\alpha} \pm i \frac{\alpha}{x} K_{i\alpha} = - K_{i\alpha \mp 1}
$$
implies
$$
{\cal E}(x) = K_{i\alpha -1}(x)K_{i\alpha +1}(x) + K_{i\alpha}^2(x).
$$
Using Nicholson's formula (Watson\cite{wat}, \S 13.72)
\begin{equation}
K_{\mu}(x) K_{\nu}(x)= 2 \int_0^{\infty} K_{\mu \mp \nu}(2 x \cosh t) \cosh (\mu \pm \nu)t~dt,
\end{equation}
it follows that
\begin{equation}
{\cal E}(x)= 2 \int_0^{\infty} [K_0 (2 x \cosh t) +K_2 (2 x \cosh
t)]\cos(2\alpha t)~dt,
\end{equation}
and since $K_0 + K_2 = -2K_1'$,
\begin{equation}
I \equiv \int_{\epsilon}^{\infty} {\cal E}(x)~dx =   \int_0^{\infty} \frac{\cos (2 \alpha  t)}{\cosh t} K_1 (2\epsilon \cosh t) dt.
\end{equation}
Since $K_1(z) \approx z^{-1}$ for $z \ll 1$, we finally obtain
\begin{equation}
I \approx   \frac{\pi \alpha}{\epsilon\sinh (\pi \alpha)}~ \label{I}
\end{equation}
as an approximation valid for $\epsilon \ll 1$.

\end{document}